\documentclass[fleqn,usenatbib]{mnras}
\usepackage[T1]{fontenc}
\usepackage{ae,aecompl}
\usepackage{graphicx}
\usepackage{needspace}
\usepackage{amssymb}
\newcommand{\psr}{PSR\,J2051$-$0827}

\title[Light curves of the companion to \psr]{Multi-colour optical light curves
  of the companion star to the millisecond pulsar \psr}

\author[V. S. Dhillon et al.]
{V. S. Dhillon,$^{1,2}$\thanks{E-mail: vik.dhillon@sheffield.ac.uk}
M. R. Kennedy,$^{3,4}$
R. P. Breton,$^{4}$
C. J. Clark,$^{4,5,6}$
\newauthor
D. Mata S\'{a}nchez,$^{2,4,7}$
G. Voisin,$^{4,8}$
E. Breedt,$^{9}$
A. J. Brown,$^{1}$
M. J. Dyer,$^{1}$
\newauthor
M. J. Green,$^{10}$
P. Kerry,$^{1}$
S. P. Littlefair,$^{1}$
T. R. Marsh,$^{11}$
S. G. Parsons,$^{1}$
\newauthor
I. Pelisoli,$^{11}$
D. I. Sahman,$^{1}$
J. F. Wild,$^{1}$
M. H. van Kerkwijk,$^{12}$
B. W. Stappers$^{4}$
\\
$^{1}$Department of Physics and Astronomy, University of Sheffield, Sheffield S3 7RH, UK \\
$^{2}$Instituto de Astrof\'{i}sica de Canarias, E-38205 La Laguna, Tenerife, Spain \\
$^{3}$Department of Physics, University College Cork, Cork, Ireland \\
$^{4}$Jodrell Bank Centre for Astrophysics, University of Manchester, Manchester M13 9PL, UK \\
$^{5}$Max Planck Institute for Gravitational Physics (Albert Einstein Institute), D-30167 Hannover, Germany \\
$^{6}$Leibniz Universit\"{a}t Hannover, D-30167 Hannover, Germany \\
$^{7}$Departamento de Astrof\'{i}sica, Universidad de La Laguna, E-38206 La Laguna, Tenerife, Spain \\
$^{8}$LUTHL, Observatoire de Paris, PSL Research University, 92195, Meudon, France \\
$^{9}$Institute of Astronomy, University of Cambridge, Cambridge CB3 0HA, UK \\
$^{10}$Department of Astrophysics, Tel Aviv University, Tel Aviv 6997801, Israel \\
$^{11}$Department of Physics, University of Warwick, Coventry CV4 7AL, UK \\
$^{12}$Department of Astronomy and Astrophysics, University of Toronto, Toronto, ON M5S 3H4, Canada \\
}

\date{\today}

\pubyear{2022}

\begin{document}
\label{firstpage}
\pagerange{\pageref{firstpage}--\pageref{lastpage}}
\maketitle

\begin{abstract}
We present simultaneous, multi-colour optical light curves of the companion star to the black-widow pulsar \psr, obtained approximately 10 years apart using ULTRACAM and HiPERCAM, respectively. The ULTRACAM light curves confirm the previously reported asymmetry in which the leading hemisphere of the companion star appears to be brighter than the trailing hemisphere. The HiPERCAM light curves, however, do not show this asymmetry, demonstrating that whatever mechanism is responsible for it varies on timescales of a decade or less. We fit the symmetrical HiPERCAM light curves with a direct-heating model to derive the system parameters, finding an orbital inclination of $55.9^{+4.8}_{-4.1}$ degrees, in good agreement with radio-eclipse constraints. We find that approximately half of the pulsar's spin-down energy is converted to optical luminosity, resulting in temperatures ranging from approximately $5150^{+190}_{-190}$\,K on the day side to $2750^{+130}_{-150}$\,K on the night side of the companion star. The companion star is close to filling its Roche lobe ($f_{\rm RL} =0.88^{+0.02}_{-0.02}$) and has a mass of $0.039^{+0.010}_{-0.011}$\,M$_{\odot}$, giving a mean density of $20.24^{+0.59}_{-0.44}$\,g\,cm$^{-3}$ and an apsidal motion constant in the range $0.0036 < k_2 < 0.0047$. The companion mass and mean density values are consistent with those of brown dwarfs, but the apsidal motion constant implies a significantly more centrally-condensed internal structure than is typical for such objects.
\end{abstract}

\begin{keywords}
stars: neutron -- pulsars: individual: \psr.
\end{keywords}

\section{Introduction}

\begin{table*}
\centering
	\caption{Journal of ULTRACAM and HiPERCAM observations of \psr. $n_{\rm exp}$ is the number of exposures and $t_{\rm exp}$ the exposure time of each frame in seconds.}
	\begin{tabular}{lcccrlcl}
		\hline
Instrument+telescope & Date start & UTC start & UTC end & $n_{\rm exp}$ & $t_{\rm exp}$ & Filters & Moon/transparency/$i$-band seeing \\
		\hline\hline
ULTRACAM+WHT & 2011/08/26 & 21:01 & 03:03 & 1085 & 20.0 & $u^{\prime}g^{\prime}i^{\prime}$ & Dark/non-photometric/0.9$^{\prime\prime}$  \\

ULTRACAM+WHT & 2011/08/27 & 20:50 & 03:27 & 1187 & 20.0~ & $u^{\prime}g^{\prime}i^{\prime}$ & Dark/photometric/1.4$^{\prime\prime}$ \\

HiPERCAM+GTC & 2021/08/06 & 22:52 & 01:26 &  300$^*$ & 30.8$^*$ & $u_{\rm s}g_{\rm s}r_{\rm s}i_{\rm s}z_{\rm s}$ & Dark/photometric/0.8$^{\prime\prime}$ \\
		\hline
    \end{tabular}
	\label{tab:journal}
$^*$\footnotesize Half the number of frames, each of double the exposure time, were obtained in $u_{\rm s}$.
\end{table*}

\begin{table}
\centering
	\caption{Magnitudes of \psr\ at light-curve maximum and minimum, measured with HiPERCAM -- see Section~\ref{sec:lc} for details.}
	\begin{tabular}{ccc}
		\hline
Filter & \multicolumn{2}{c}{Magnitude at} \\
& max & min \\
		\hline\hline
$u_{\rm s}$ & $25.0\pm0.1$   & $>25.9$ \\
$g_{\rm s}$ & $23.47\pm0.02$ & $>26.8$ \\
$r_{\rm s}$ & $22.67\pm0.01$ & $27.9\pm0.4$ \\
$i_{\rm s}$ & $22.30\pm0.01$ & $25.6\pm0.1$ \\
$z_{\rm s}$ & $22.12\pm0.02$ & $24.8\pm0.2$ \\
		\hline
    \end{tabular}
	\label{tab:mags}
\end{table}

Pulsars are highly-magnetised, rotating neutron stars. Over 3000 are
known\footnote{http://www.atnf.csiro.au/research/pulsar/psrcat}, most
of which have spin periods in the 0.1-1\,s range that gradually
increase with time due primarily to the emission of magnetic dipole
radiation. After tens of millions of years, the spin slows to such an
extent that the mechanism powering the radio emission turns off and
the pulsar `dies'. There exist a sub-set of more than 500 known
pulsars\footnote{http://astro.phys.wvu.edu/GalacticMSPs/GalacticMSPs.txt} \citep{Manchester2005+ATNF},
however, that have the fastest spin periods, of order milliseconds,
and are believed to be much older ($\sim10^{9}$\,yr) than ordinary
pulsars. These so-called `millisecond pulsars' (MSPs) are believed
to be dead pulsars that have been spun up (or `recycled') by the
accretion of mass from a companion star via Roche-lobe
overflow (see \cite{tauris06} and references therein). 
During this accretion phase, the object appears as an X-ray binary. Once accretion has stopped, the pulsar begins emitting in the radio again. In some of the closer binaries, particle and $\gamma$ radiation from the pulsar is then believed to ablate the companion star, possibly evaporating it entirely to leave an isolated millisecond pulsar (see \cite{polzin20} and references therein). Hence such systems are sometimes referred to as `black-widow pulsars'. They consist of a
millisecond pulsar in a tight orbit ($P \lesssim 24$\,h) with a very low mass
companion star ($M_{2} \lesssim 0.05 M_{\odot}$) and usually exhibit radio eclipses each orbit due to the obscuration of the pulsar by the ablated material. Objects in each of the evolutionary phases described above have been discovered, which lends support to this general picture, although many uncertainties remain. For a review of MSPs, see \cite{lorimer08}.

\psr\ was the second black-widow pulsar to be discovered in the Galactic disk \citep{stappers96a}, after PSR\,J1959+2048 (also known as PSR\,B1957+20; \citealt{fruchter88}), and is a 4.5\,ms pulsar in a 2.4\,hr period orbit with a low mass companion. The companion star was detected in the optical by \cite{stappers96b}, and orbital light curves were subsequently obtained by \cite{stappers99} and \cite{stappers01}. The nature of the companion star in the \psr\ system remains uncertain, due to the fact that asymmetries were observed in the light curves of \cite{stappers01} which resulted in fits that were unable to distinguish between companion stars that were almost filling their Roche lobes and those that were only half filling their Roche lobes. Not knowing the companion star radius makes it difficult to determine if the companion star is a white dwarf, a brown dwarf
or a semi-degenerate helium star \citep{lazaridis11}. It also makes it difficult to predict the future evolution of the system -- if the companion star is close to Roche-lobe filling, much less pulsar energy is required to drive a wind from its surface, and mass may also be lost through the inner-Lagrangian point, hastening the evaporation timescale to form an isolated millisecond pulsar (e.g. \citealt{levinson91}). Knowing the Roche-lobe filling factor also allows the apsidal motion constant of the companion to be determined from the detection of  its gravitational quadrupole moment by \cite{voisin20}, enabling the internal structure of the star to be modelled. 

In this paper, we present new simultaneous multi-colour light curves of \psr\  and use them to determine the system parameters and the nature of the companion star. 

\section{Observations and data reduction}

In 2011, we observed \psr\ simultaneously in $u^{\prime}g^{\prime}i^{\prime}$ using the high-speed, triple-beam camera ULTRACAM \citep{dhillon07} on the 4.2\,m William Herschel Telescope (WHT) on La Palma. In 2021, we  observed \psr\ again, this time simultaneously in $u_{\rm s}g_{\rm s}r_{\rm s}i_{\rm s}z_{\rm s}$ using the high-speed, quintuple-beam camera HiPERCAM \citep{dhillon21} on the 10.4\,m Gran Telescopio Canarias (GTC) on La Palma. Note that both filter sets use the same cut points as the original Sloan Digital Sky Survey (SDSS) filters \citep{fukugita96}, and the primed filters used by ULTRACAM also have similar bandpass shapes to SDSS, but the HiPERCAM subscript-s filters are more top-hat in shape and have significantly higher throughputs, particularly in the $u_{\rm s}$ and
$g_{\rm s}$ bands (see \citealt{dhillon21}). A journal of observations is presented in Table~\ref{tab:journal}. Both instruments
were used in full-frame, no-clear mode, giving a dead time between each frame in HiPERCAM and ULTRACAM of 0.008\,s and 0.024\,s, respectively, where each 
HiPERCAM/ULTRACAM frame is GPS time-stamped to a relative (i.e. frame-to-frame) accuracy of 0.1/50\,$\mu$s and an absolute accuracy of 0.1/1\,ms, respectively (\citealt{dhillon07}, \citeyear{dhillon21}).

The ULTRACAM and HiPERCAM data were reduced using their respective data reduction pipelines (\citealt{dhillon07}, \citeyear{dhillon21}). All frames were debiased and then flat-fielded, the latter using the median of twilight-sky frames taken with the telescope spiralling. The CCD fringing pattern was removed from the $z_{\rm s}$ HiPERCAM frames using the median of night-sky frames taken with the telescope spiralling. \psr\ was
invisible in $u^{\prime}$ in the ULTRACAM data, so this band shall not be discussed further, and we skipped the readout of every other $u_{\rm s}$ HiPERCAM frame using the {\em NSKIP} parameter (see \citealt{dhillon21}) in order to double the exposure time
(and halve the number of frames) in this band.

We used software apertures that scaled in size with the seeing to extract the counts from \psr\ and a number of comparison stars in the same field of view, the latter acting as the reference for the PSF fits, transparency and extinction corrections. The comparison stars were also used for flux calibration via their magnitudes given in the Pan-STARRS1 catalogue (see~\citealt{magnier20}~and~references~therein)
and converted to SDSS~magnitudes \citep{finkbeiner16}. The aperture position of \psr\ relative to a bright comparison star was determined from a sum of all the images, and this offset was then held fixed during the reduction so as to avoid aperture centroiding problems during light-curve minimum. The effect of atmospheric refraction on the relative aperture positions is negligible due to the similarity in colour between the target and our chosen reference star, and the fact that our observations on each night were approximately centred on meridian transit. The sky level was determined from a clipped mean of the counts in an annulus surrounding each star and subtracted from the object counts.

\section{Results}

\subsection{Light curves}
\label{sec:lc}

\begin{figure*}
\includegraphics[clip, trim=1.8cm 1.0cm 2.9cm 0.7cm, width=0.9\textwidth]{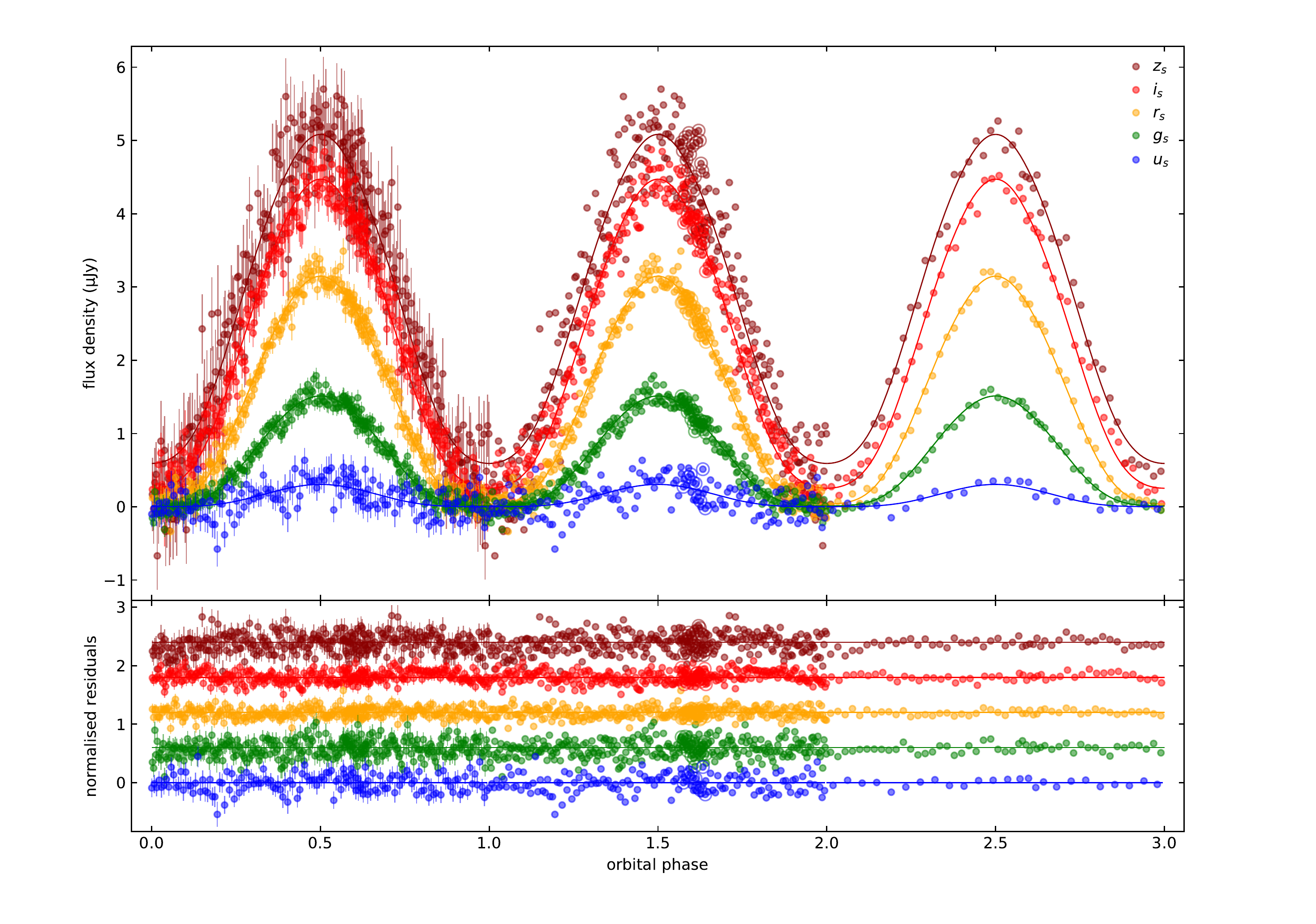}\newline
\caption{Upper panel: HiPERCAM $u_{\rm s}g_{\rm s}r_{\rm s}i_{\rm s}z_{\rm s}$ light curves (from bottom to top, the blue/green/orange/red/maroon points, respectively) of \psr, phased according to the ephemeris given in Eqn.~\ref{eqn:ephemeris}. The folded light curve has been repeated three times, the first cycle with error bars, the second without, and the third cycle with every 6 data points binned, giving an effective exposure time of approximately 3\,min per point. The solid lines show the {\sc icarus} fits to the HiPERCAM data described in Section~\ref{sec:modelling}. The HiPERCAM data cover approximately 1.08 binary orbits -- the ringed points around phase 1.6 show the overlapping data from the second orbit, demonstrating the excellent agreement with the data taken during the first orbit. Lower panel: Normalised residuals of the {\sc icarus} fits to the HiPERCAM data, offset vertically from each other by 0.6 and with the $u_{\rm s}$ residuals divided by a factor of 10 for clarity.}
\label{fig:lc_hcam}
\end{figure*}

\begin{figure*}
\includegraphics[clip, trim=1.8cm 1.0cm 2.8cm 0.7cm, width=0.9\textwidth]{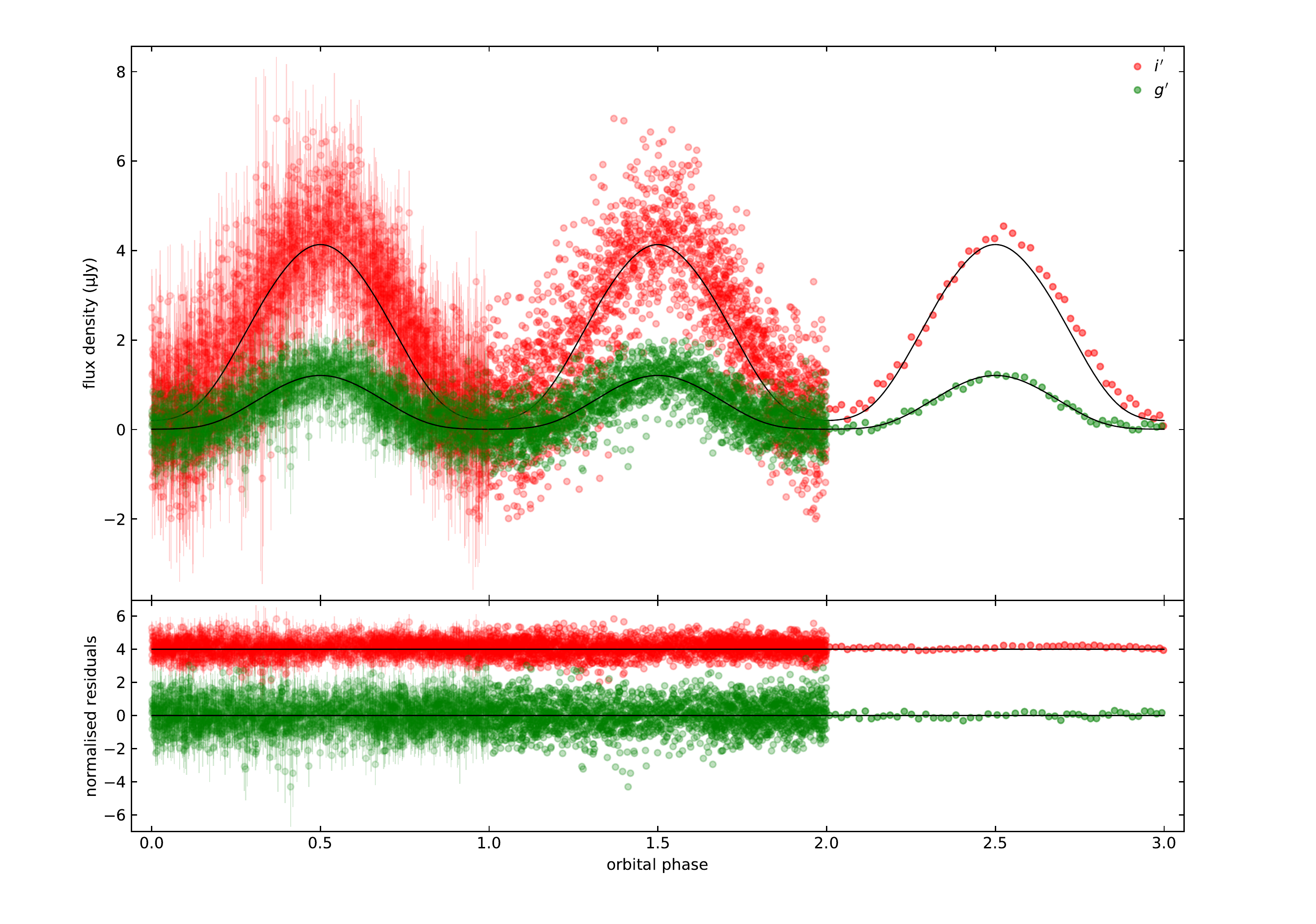}
\caption{Upper panel: ULTRACAM $g^{\prime}i^{\prime}$ light curves (green (bottom) and red (top) points, respectively) of \psr, phased according to the ephemeris given in Eqn.~\ref{eqn:ephemeris}. The folded light curve has been repeated three times, the first cycle with error bars, the second without, and the third cycle with every 45 data points binned, giving an effective exposure time of approximately 15\,min per point. The solid lines show the result of fitting the HiPERCAM model presented in Fig.~\ref{fig:lc_hcam} to the ULTRACAM data, allowing only the flux scaling factors to vary. Lower panel: Normalised residuals of the {\sc icarus} fits to the ULTRACAM data, offset vertically from each other by 4.0.}
\label{fig:lc_ucam}
\end{figure*}

\psr\ was the subject of a 21\,yr radio timing study by \cite{shaifullah16}. Since then, they have continued to monitor \psr\ and provided us with the following
up-to-date binary ephemeris:

\begin{equation}
\begin{array}{lrll}
T_{\rm asc} = & \hspace*{-0.2cm} {\rm BMJD}\ 59099.9673395
& \hspace*{-0.2cm}+\ 0.09911025846 & \hspace*{-0.3cm} E \\
& \pm\ 0.0000019 & \hspace*{-0.2cm} \pm\ 0.00000000014, & \\
\label{eqn:ephemeris}
\end{array}
\end{equation}
where BMJD refers to the Modified Julian Date on the Barycentric Dynamical Timescale (TDB) and $E$ is the cycle number. This ephemeris is stable and precise enough to be applicable to both our HiPERCAM observations in 2021 and our ULTRACAM observations in 2011. The HiPERCAM and ULTRACAM light curves we obtained of \psr, folded on the above ephemeris, are shown in Figs.~\ref{fig:lc_hcam} and~\ref{fig:lc_ucam}, respectively. Note that $T_{\rm asc}$ in Eqn.~\ref{eqn:ephemeris} corresponds to the epoch of the pulsar's ascending node. In what follows, we have chosen to apply a phase offset of $-0.25$ so that phase 0 corresponds to the superior conjunction of the pulsar.

The light curve of \psr\ exhibits a single, broad hump that is brightest at phase 0.5 and faintest at phase 0, indicating that the dominant source of optical emission is the irradiated inner hemisphere of the companion star; the neutron star itself is invisible at optical wavelengths in all (non-transitional) MSPs\footnote{Optical pulsations that may originate directly from the pulsar have been observed in the transitional system PSR J1023+0038 \citep{ambrosino17}.}. The $g_{\rm s}$ and $r_{\rm s}$ light curves show evidence for a possible flaring event just prior to phase 0.5 --  such flaring activity has been seen before in black-widow systems, e.g. \citet{romani12}. The $i^{\prime}$-band ULTRACAM light curve (Fig.~\ref{fig:lc_ucam}) shows evidence for an asymmetry, confirming the finding of \cite{stappers01}, and suggesting that the leading edge of the companion star is brighter than the trailing edge, leading to a flux excess after the peak. At first glance, the $g^{\prime}$-band ULTRACAM light curve does not appear to show this asymmetry, but it is also present at a lower
level (see Section~\ref{sec:asymmetry}). The HiPERCAM light curves obtained a decade later (Fig.~\ref{fig:lc_hcam}), on
the other hand, are symmetrical. For reference, the magnitudes at light-curve maximum/minimum of \psr\ are given in Table~\ref{tab:mags}. These were determined from a clipped-mean of the HiPERCAM fluxes in the phase ranges 0.45--0.55 at light-curve maximum and 0.9--1.1 at minimum -- note there was no significant flux detected at light-curve minimum in $g_{\rm s}$ and $u_{\rm s}$ and so the measured 5$\sigma$ limiting magnitudes are quoted instead.

%
%
%
%

\subsection{Light-curve modelling}
\label{sec:modelling}

The multi-colour light curves of \psr\ were fit using the {\sc icarus} modelling software \citep{breton12}. The companion star is assumed to be tidally locked to the pulsar and its surface is modelled as a finite-element grid, with the intensity of each element calculated from a BT-Settl\footnote{https://phoenix.ens-lyon.fr/Grids/} {\sc phoenix} model atmosphere \citep{allard14} appropriate to the physical properties (temperature, gravity, velocity) at that location on the stellar surface. The model atmospheres are folded through the HiPERCAM and ULTRACAM filter profiles and the observed flux in each photometric band is then obtained by integrating the specific intensity emerging from the surface and visible to an observer located at a given direction and distance. The best-fit model parameters and their errors are determined using the MultiNest nested sampling algorithm \citep{feroz13}, as implemented in the Python package {\sc PyMultiNest} \citep{buchner14}. 

The input parameters are: the ephemeris ($T_{\rm asc}$, $P$) given in Eqn.~\ref{eqn:ephemeris}; the projected semi-major axis of the pulsar orbit, $x= a_{\rm psr} \sin i = 0.04507837(398)$ lt-s, 
from the radio-timing data described in Section~\ref{sec:lc}; the gravity darkening exponent, $\beta$. For the latter, we assumed a value of $\beta=0.08$, appropriate for stars with convective envelopes, a reasonable assumption for the cool, low-mass companion in \psr\ \citep{stappers01}. As a check on the robustness of our fit parameters to this assumption, we also modelled the light curves using $\beta=0.25$, a value more appropriate for stars with radiative envelopes \citep{lucy67}, and found negligible ($< 1\sigma$) differences in the resulting parameters. 

The fit parameters are as follows:

\begin{itemize}
\item $E(g-r)$ -- the interstellar reddening. We adopted a Gaussian prior of $E(g-r) = 0.10 \pm 0.02$, measured in the direction of \psr\ from the 3D dust maps of \cite{green19}\footnote{http://argonaut.skymaps.info}, which is valid for $d>0.73$\,kpc. The extinction in each band, $A$, is then calculated from the extinction vectors, $R$, given by \cite{green19}.

\item $d$ -- the distance to the binary. No {\em Gaia} or radio timing parallax is available for \psr, so we
adopted a prior based on the dispersion measure, $DM=20.7299$\,pc\,cm$^{-3}$ \citep{shaifullah16}, which corresponds to a distance of 1.469\,kpc using the Galactic free-electron density model of \cite{yao17}\footnote{https://www.atnf.csiro.au/research/pulsar/ymw16}. We adopted
a log-normal prior on this DM-derived distance, with a fractional error of 0.45 \citep{yao17}. Following \cite{clark21}, we multiplied the DM-distance prior by two additional priors. First, we adopted a prior based on the \cite{levin13} model for the density of MSPs in the Galactic disk, which has a Gaussian profile in Galactic radius with width $\sigma = 4.5$\,kpc, and an exponential profile in height above the Galactic plane with scale height $z = 0.5$\,kpc. Second, we adopted a prior based on the transverse velocities of binary MSPs, which can be approximated by an exponential distribution with a mean value of $93\pm13$\,km\,s$^{-1}$ \citep{desvignes16}. The radio timing of \psr\ by \cite{shaifullah16} gives a total proper motion of $\mu = 6.1\pm0.1$\,mas\,yr$^{-1}$, from which
the transverse velocity can be calculated: $v_{\rm T} = 28.9\,d$\,km\,s$^{-1}$, where $d$ is in kpc. We multiplied the distance prior by $e^{-v_{\rm T}} / 93$ to take this into account.

\item $T_{\rm base}$ -- the base temperature of the surface of the companion
star, defined as the temperature at the pole of the star prior to irradiation. To account for gravity darkening, the base temperature at each point on the stellar surface is multiplied by the factor $(g/g_{\rm pole})^{\beta}$, where $g$ is the acceleration due to gravity. $T_{\rm base}$ was constrained to lie within the range $2100 - 5000$\,K. The lower limit of the model atmospheres we used is 2300\,K; we verified that extrapolating them to 2100\,K introduces no significant systematic errors.

\item $T_{\rm irr}$ -- the irradiating temperature, which accounts for the 
effect of heating by the pulsar. $T_{\rm irr}$ is defined with respect to
the centre of mass of the companion star at a distance $a$ from the
pulsar, which would receive a flux of $\sigma T_{\rm irr}^{4}$, where $\sigma$ is the Stefan-Boltzmann constant. Hence a location on the stellar surface at a distance $r$ from the pulsar, and which has a normal vector at an angle $\theta$ from the vector pointing to the pulsar, receives heating power per unit area of $\sigma T_{\rm irr}^{4} \cos \theta\,a^{2}/r^{2}$. Assuming that the pulsar's irradiating flux is immediately thermalised and re-radiated \citep{breton13},
the surface temperature at this location is raised to 
$T = [T_{\rm base}^{4}(g/g_{\rm pole})^{4\beta} + T_{\rm irr}^{4} \cos \theta\,a^{2}/r^{2}]^{1/4}$. We also investigated the 
approach of \cite{romani16}, who applied gravity darkening after
rather than before irradiation, and found that it did not alter the main conclusions of this paper.

\item $f_{\rm RL}$ -- the Roche-lobe filling factor, defined as the ratio between the companion's radius in the direction towards the pulsar and the distance between the companion's centre of mass and inner-Lagrangian point, 
$L_{\rm 1}$. A uniform prior was adopted, limiting the value between $0.2<f_{\rm RL}<1$.

\item $i$ -- the binary inclination angle. A prior that was uniform in $\cos i$ was adopted to allow for the fact that high inclinations are more likely to be observed if the binary orbits are randomly oriented.

\item $M_{\rm psr}$ -- the mass of the pulsar. A uniform prior was adopted, limiting the value between $1.0<M_{\rm psr}<2.5$\,M$_{\odot}$
(comfortably encompassing all known, reliable neutron star masses\footnote{e.g. https://www3.mpifr-bonn.mpg.de/staff/pfreire/NS\_masses.html}).
The $M_{\rm psr}$ and $i$ fit parameters, in conjunction with the $P$ and $x$ input parameters, were used to derive 
the mass ratio ($q=M_{\rm psr}/M_{\rm c}$) from the binary mass function:
\begin{equation}
\frac{M_{\rm psr} \sin^3 i}{(1 + 1/q)^2} = \frac{K_{\rm c}^3 P}{2 \pi G} = \frac{q^3 x^3 4 \pi^2}{G P^2},
\label{eqn:massfn}
\end{equation}
which in turn allowed the companion mass $M_{\rm c}$ and companion radial velocity $K_{\rm c}$ to be derived using the above relations. The light curve constrains
$i$, and we chose to fit $M_{\rm psr}$ rather than $q$ or $K_{\rm c}$ as we can place a more informative prior on $M_{\rm psr}$ based on our knowledge of the observed neutron star mass distribution than we can on the relatively unconstrained $q$ and $K_{\rm c}$.

\end{itemize}

As well as deriving $q$, $M_{\rm c}$ and $K_{\rm c}$ from the fit parameters, we also used them to derive the volume-averaged radius, $R_{\rm c}$, and density, $\rho_{\rm c} = M_{\rm c}/\frac{4}{3}\pi R_{\rm c}^3$, of the companion star, and the heating efficiency $\varepsilon = L_{\rm irr}/\dot{E}$. The latter compares the luminosity of the pulsar, $L_{\rm irr} = 4\pi a^2 \sigma T_{\rm irr}^4$, where $a=x(1+q)/\sin{i}$ is the orbital separation, to the pulsar's spin-down power, $\dot E = 4\pi^2 I \dot{P} / P^3$, assuming a canonical value for the neutron-star moment of inertia of $I={\rm 10}^{45}$\,g\,cm$^2$ (e.g. \citealt{abdo13}) and values for the pulsar spin period and its derivative from the radio-timing data described in Section~\ref{sec:lc}.

At each step in the sampling process, the input and fit parameters described above are used to calculate the fluxes received on Earth from the modelled companion star in each filter. To account for any systematic errors in the flux calibration, extinction and atmosphere models, we allowed a flux scaling factor (or equivalently, a magnitude offset) in each band, penalized by a zero-mean Gaussian prior with a width of 0.1 mag.
The latter value represents the uncertainty in our flux calibration given that we did not account for the colour terms relating to the differences between and within the HiPERCAM Super-SDSS and ULTRACAM SDSS filter systems compared to SDSS (for a detailed discussion of these differences, see \citealt{brown22} and \citealt{wild22}, respectively). Without such a prior, we would obtain unrealistically small errors in the distance and reddening due to the degeneracy between these parameters and the flux scaling factor. For reasons that will become apparent below, we chose to fit the HiPERCAM $u_{\rm s}g_{\rm s}r_{\rm s}i_{\rm s}z_{\rm s}$ and ULTRACAM $g^{\prime}i^{\prime}$ data separately, but for each instrument we fit all of the  filters simultaneously.

\begin{figure*}
\centering
\includegraphics[width=17.5cm,angle=0]{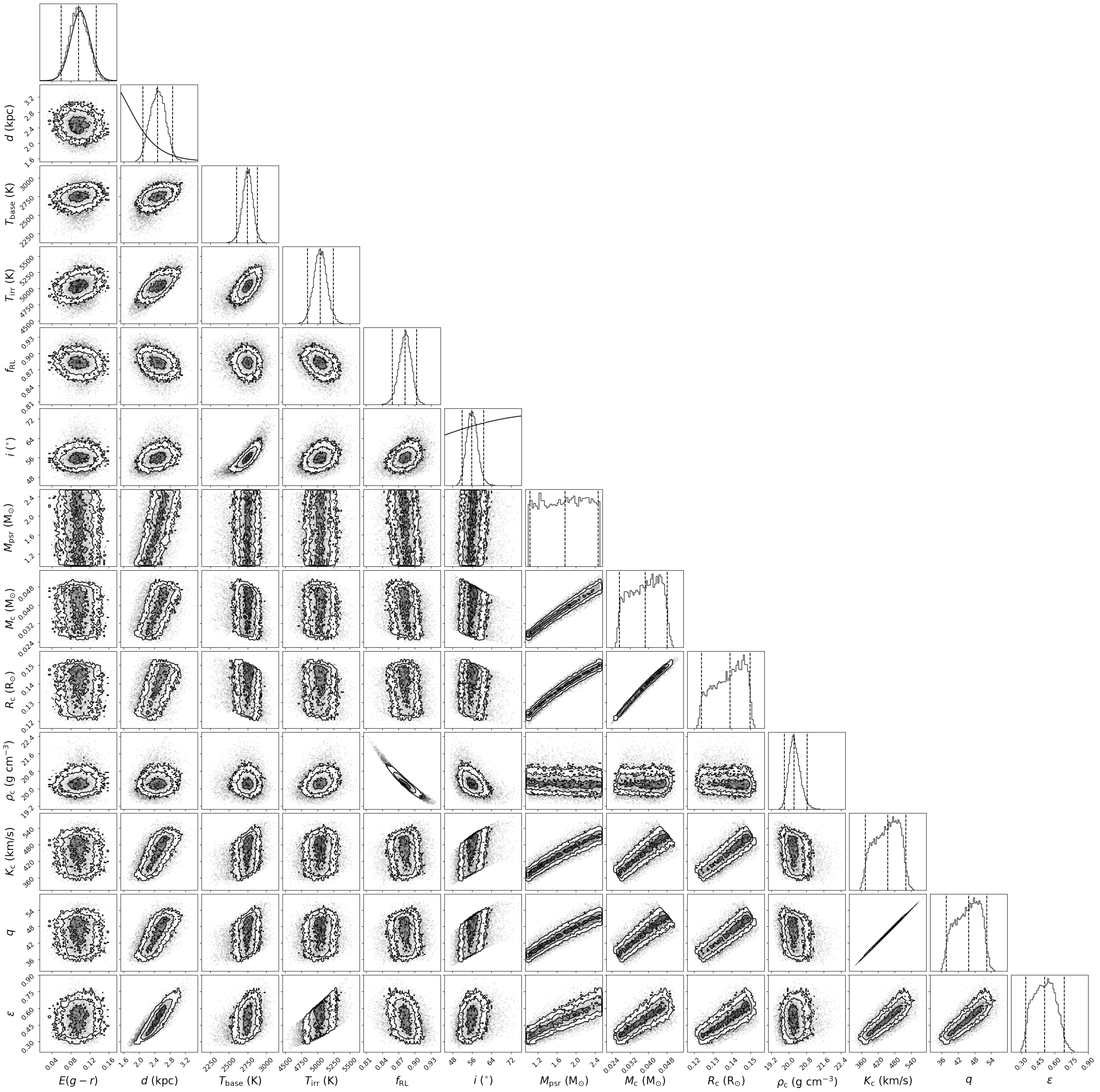}
\caption{Corner plot showing the posterior distributions of the light-curve fit parameters. The last six parameters ($M_{\rm c}$, $R_{\rm c}$, $\rho_{\rm c}$, $K_{\rm c}$, $q$, $\varepsilon$) were derived from the values of the fit parameters and the input parameters. The contours indicate the 1$\sigma$, 2$\sigma$ and 3$\sigma$ confidence regions. The dashed vertical lines on the histograms show the mean and 95\%\ confidence intervals of the parameter distributions, and where non-uniform priors were used, these are shown as curves on the histograms.}
\label{fig:corner}
\end{figure*}

The {\sc icarus} fits to the HiPERCAM $u_{\rm s}g_{\rm s}r_{\rm s}i_{\rm s}z_{\rm s}$ light curves are shown as the solid curves in the upper panel of Fig.~\ref{fig:lc_hcam}, the resulting fit and derived parameters are given in Table~\ref{tab:fitparams}, and the posterior distributions of these parameters are shown in Fig.~\ref{fig:corner}. The latter figure shows that, of the three parameters with non-uniform priors, only the prior on $E(g-r)$ has a noticeable effect on the posterior distributions compared to a uniform prior. $E(g-r)$ is essentially unconstrained by the data, and so the posterior distribution of this parameter closely follows the prior, but $E(g-r)$ is slightly correlated with $T_{\rm irr}$ and $T_{\rm base}$, and so the effect of the Gaussian prior on this parameter is to slightly reduce the uncertainties on these temperatures. The posterior distributions of $i$ and $d$, on the other hand, are much narrower than the priors, indicating that their inferred values are dominated by the data and not by the priors. Judging from the fit residuals shown in the lower panel of Fig.~\ref{fig:lc_hcam}, the reduced-$\chi^2$ value of 0.96, the near-unity flux scaling factors given in Table~\ref{tab:fitparams} (which all lie well within the 0.1 mag Gaussian priors we set), and the shape of the posterior distributions in Fig.~\ref{fig:corner}, the fit to the HiPERCAM data is acceptable, and the resulting parameter values will be discussed in Section~\ref{sec:sysparams}.

\renewcommand*\arraystretch{1.4}
\begin{table}
	\centering
	\caption{Results of the {\sc icarus} fits to the HiPERCAM $u_{\rm s}g_{\rm s}r_{\rm s}i_{\rm s}z_{\rm s}$ light curves. The parameter values quoted are the median of the marginalised posterior distributions shown in Fig.~\ref{fig:corner}, with the 95\% confidence region given in sub- and superscript.}
	\begin{tabular}{lc}
		\hline
		\multicolumn{2}{c}{$\chi^2$/degrees of freedom = 1287.4/1341} \\
		\multicolumn{2}{c}{$u_{\rm s}g_{\rm s}r_{\rm s}i_{\rm s}z_{\rm s}$ flux scaling factors = 1.06/0.98/0.96/0.98/0.94} \\
		\hline
		\multicolumn{2}{c}{Fit parameters} \\
		\hline
$E(g-r)$   & $0.096^{+0.038}_{-0.037}$ \\
$d$ (kpc)                    & $2.48^{+0.39}_{-0.38}$ \\
$T_{\rm base}$ (K)              & $2750^{+130}_{-150}$ \\
$T_{\rm irr}$ (K)            & $5040^{+210}_{-200}$ \\
$f_{\rm RL}$                 & $0.88^{+0.02}_{-0.02}$ \\
$i$ ($^{\circ}$)             & $55.9^{+4.8}_{-4.1}$ \\
$M_{\rm psr}$ (M$_{\odot}$)  & $1.77^{+0.69}_{-0.73}$ \\
		\hline
		\multicolumn{2}{c}{Derived parameters} \\
		\hline
$M_{\rm c}$ (M$_{\odot}$)         & $0.039^{+0.010}_{-0.011}$ \\
$R_{\rm c}$ (R$_{\odot}$)         & $0.139^{+0.011}_{-0.015}$ \\
$\rho_{\rm c}$ (g\,cm$^{-3}$)           & $20.24^{+0.59}_{-0.44}$ \\
$K_{\rm c}$ (km\,s$^{-1}$)              & $454^{+66}_{-81}$ \\
$q \equiv M_{\rm psr}/M_{\rm c}$  & $45.8^{+6.7}_{-8.2}$ \\
$\varepsilon$                     & $0.51^{+0.18}_{-0.17}$ \\
$k_2$                             & $0.0036-0.0047$ \\
		\hline
	\end{tabular}
	\label{tab:fitparams}
\end{table}


However, the same can not be said for the {\sc icarus} fits to the ULTRACAM $g^{\prime}i^{\prime}$ data. Like \cite{stappers01}, we found that the asymmetry in the light curves prevented a reliable model from being found. This is unsurprising given that the simple, direct-heating model described above can only produce symmetric light curves. So we experimented with different modifications to the direct-heating model to fit the asymmetry: allowing for a small orbital phase offset (e.g. \citealt{nieder19}); using spherical harmonics or hot and cold and spots to account for brightness asymmetries on the stellar surface (e.g. \citealt{clark21}); allowing for heat redistribution due to convective flows on the stellar surface (e.g. \citealt{voisin20}, \citealt{stringer21}). Although acceptable fits could be obtained by adding free parameters in this way, the various methods did not point to a consistent set of fit parameters: in particular, the best-fit inclinations and Roche-lobe filling factors varied widely between models.

Given that we do not know which of the above methods of introducing light-curve asymmetries in \psr\ is physically correct, and that the HiPERCAM data are of higher quality than the ULTRACAM data and do not show the asymmetry, we decided to disregard the ULTRACAM fits. Instead, we assume that the HiPERCAM light curve represents the underlying system (i.e. no asymmetric heating or spots), and we adopt the HiPERCAM fit as correctly representing the binary parameters. For reference, we took the best-fit model to the HiPERCAM light curve and fit it to the ULTRACAM data, allowing only the flux scaling factors to vary. 
The resulting fit to the ULTRACAM data ($\chi^2=5707$, degrees of freedom $=4492$) is shown as the solid curves in Fig.~\ref{fig:lc_ucam} and required flux scaling factors of $g^{\prime} = 0.82$ and $i^{\prime} = 0.93$, in the sense that the best-fit HiPERCAM model is multiplied by these factors to match the ULTRACAM fluxes. It can be seen that the HiPERCAM model provides a reasonable match to the $g^{\prime}$ ULTRACAM light curve, but the fit to the $i^{\prime}$ light curve is poor, particularly after phase 0.5 due to the extra flux from the leading edge of the companion star. We shall discuss the possible origins of this extra flux in Section~\ref{sec:asymmetry}.

\section{Discussion}

\subsection{System parameters}
\label{sec:sysparams}

With the exception of the Roche-lobe filling factor, which shall be discussed further below, our light-curve fit parameters are in reasonable agreement with those of \cite{stappers01}. We are able to confirm that \psr\ has only a moderate inclination of $55.9^{+4.8}_{-4.1}$ degrees, slightly higher than the value of $i\sim40^{\circ}$ found by \cite{stappers01} and more consistent with the model of plasma-lensing during radio eclipse derived by \cite{lin21}. Like \cite{stappers01}, we find that the minimum night-side temperature ($\sim T_{\rm base}$) of the companion star is cool ($2750^{+130}_{-150}$\,K). This is close to the 2300\,K minimum of our model-atmosphere grid, which could be the cause of the slight over-prediction of the $z_{\rm s}$ flux at minimum evident in Fig.~\ref{fig:lc_hcam}. In contrast, the maximum day-side temperature ($\sim [T_{\rm base}^{4}+T_{\rm irr}^{4}]^{1/4}$) is $5150^{+190}_{-190}$\,K. We find that approximately half of the pulsar's spin-down energy is converted to optical luminosity, which is somewhat higher than the value of $\sim20$\% typically observed in black-widow systems (\citealt{breton13}, \citealt{draghis19}). 

\psr\ is detected in $\gamma$-rays by the {\em Fermi} Large Area Telescope \citep{Wu2012+J2051}, which measures an integrated energy flux above 100 MeV of $F_{\gamma} = (2.5\pm0.3)\times10^{-12}$\,erg\,cm$^{-2}$\,s$^{-1}$ \citep{4FGL_DR3}. For our estimated $d=2.5 \pm 0.2$\,kpc (1$\sigma$ uncertainties), this corresponds to a luminosity of $L_{\gamma} = 4\pi F_{\gamma}d^2 = (1.9 \pm 0.4) \times10^{33}$ erg s$^{-1}$. This $\gamma$-ray emission therefore represents $\sim35$\% of the pulsar's spin-down power budget, which is at the lower limit of the range of $\epsilon$ inferred from our {\sc Icarus} model. Beaming effects may result in the $\gamma$-ray flux being stronger at the pulsar's spin equator, which should be aligned with the orbital plane since the pulsar is believed to have been spun-up by accretion, than when viewed at $i \sim 56^{\circ}$, and so the pulsar's $\gamma$-ray emission may be sufficient to explain the irradiation in this system.

Without a measurement of the radial velocity of the companion star we are unable to determine the pulsar mass, but by assuming it lies in the range $1.0<M_{\rm psr}<2.5$\,M$_{\odot}$, the light-curve fit indicates that the companion star has a mass in the range $0.039^{+0.010}_{-0.011}$\,M$_{\odot}$, similar to the values derived for other black-widow systems (\citealt{roberts13}, \citealt{draghis19}). The Roche-lobe filling factor lies in the range $0.88^{+0.02}_{-0.02}$, resolving the ambiguity noted by \cite{stappers01} in favour of a companion star that is close to filling its Roche lobe. The resulting volume-averaged radius of the companion star ($0.139^{+0.011}_{-0.015}$\,$R_{\odot}$) implies a mean density of $20.24^{+0.59}_{-0.44}$\,g\,cm$^{-3}$. 
The mass and mean density of the companion star in \psr\ are consistent with those of a hydrogen brown dwarf rather than a helium or carbon white dwarf (\citealt{tang14}, \citealt{hatzes15}, \citealt{kaplan18}).

With a reliable estimate of the Roche-lobe filling factor, we can now determine the apsidal motion constant, $k_2$, from the orbital precession measurement of \psr\ by \cite{voisin20}, constraining it to the range $0.0036 < k_2 < 0.0047$.\footnote{This value has been calculated using the mass ratio given in Table~\ref{tab:fitparams}, not by using Fig. 1 of \cite{voisin20}. Note also that the ordinate of the latter figure is incorrect due to an error of a factor of 10 in the assumed mass ratio.} The apsidal motion constant describes how centrally condensed an object is, with higher central concentrations corresponding to smaller values of $k_2$. Our measured value is two orders of magnitude smaller than those typical of brown dwarfs and the gas giants of the solar system \citep{heller10}. Detailed interpretation of our $k_2$ measurement requires integration of the equations of stellar structure and is outside the scope of this paper.

\subsection{Light-curve asymmetry}
\label{sec:asymmetry}

Asymmetric optical light curves have been seen in a number of other black-widow systems, including PSR\,J1810+1744 \citep{romani21}, PSR\,J1311$-$3430 \citep{romani15}, PSR\,J1653$-$0158 \citep{nieder20}, PSR\,J0952$-$0607 \citep{nieder19} and PSR J1959+2048 \citep{kandel20a}. Models for the origin of the asymmetry include ducting of the particles in the shock between the pulsar and companion star winds (the intra-binary shock, IBS) onto the magnetic poles of the companion star \citep{sanchez17}, re-processed radiation from the IBS \citep{romani16}, hot and cold spots on the companion star's surface (\citealt{vanstaden16}, \citealt{clark21}), and redistribution of energy on the companion star's surface by convection (\citealt{voisin20}, \citealt{kandel20a}).

\begin{figure}
\includegraphics[clip, trim=1.6cm 0.2cm 10.0cm 0.2cm, width=0.48\textwidth]{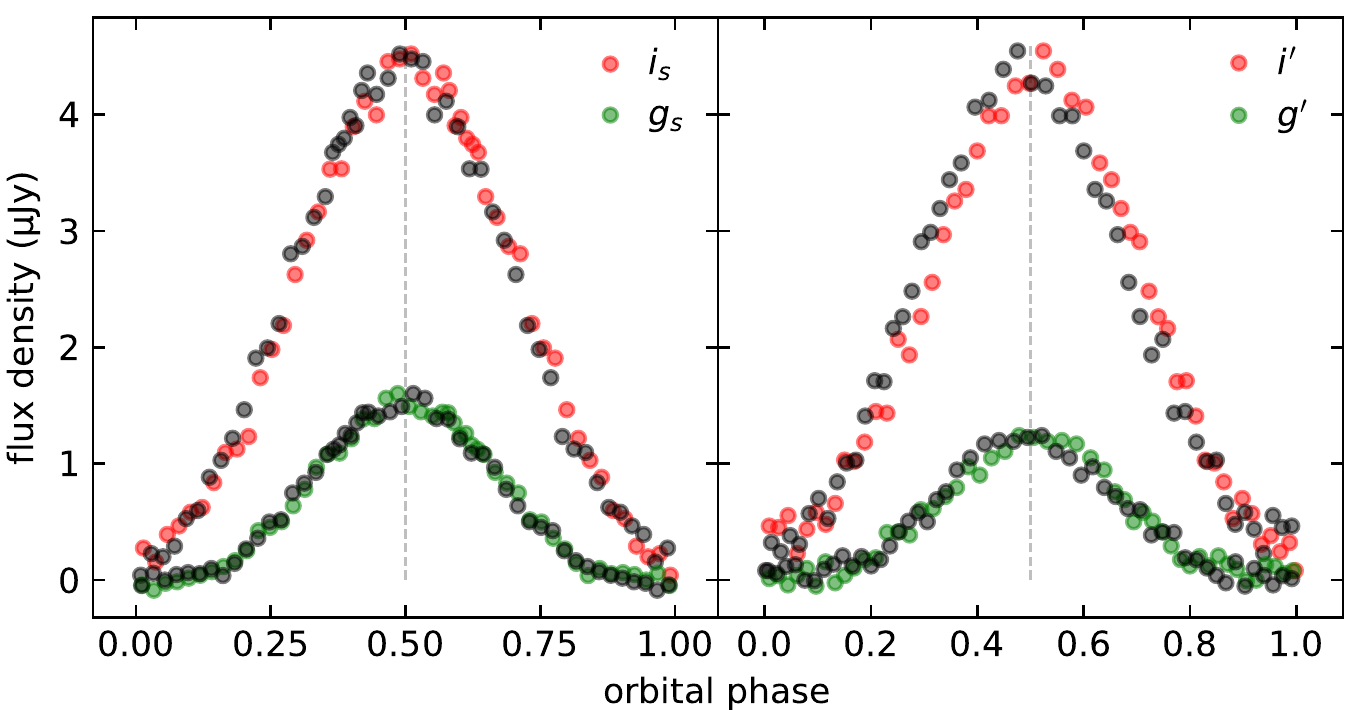}
\caption{HiPERCAM $g_{\rm s}i_{\rm s}$ (left) and ULTRACAM $g^{\prime}i^{\prime}$ (right) light curves. To emphasize any asymmetries present, the black points show the 
light curves mirrored around phase 0.5 (indicated by the
vertical dashed line).}
\label{fig:lc_asymm}
\end{figure}

The $i^{\prime}$-band ULTRACAM light curve presented in Fig.~\ref{fig:lc_ucam} is clearly asymmetric around phase 0.5. By comparing this light curve to the {\sc icarus} fit and the simultaneous $g^{\prime}$-band ULTRACAM light curve plotted in the same figure, we can infer that the source of the asymmetry is predominantly excess $i^{\prime}$-band light on the leading hemisphere of the companion star. This asymmetry appears to be absent in the HiPERCAM light curves obtained a decade later, as highlighted in Fig.~\ref{fig:lc_asymm}. A close inspection of Fig.~\ref{fig:lc_asymm} shows that the asymmetry is actually present in the $g^{\prime}$-band ULTRACAM light curve as well, but is weaker and covers a narrower phase range than the $i^{\prime}$-band asymmetry.

Similar long-term changes in the light-curve symmetry of MSP binaries have been observed before, e.g. in PSR\,J2339$-$0533 \citep{kandel20b} and PSR\,J1723$-$2837 \citep{vanstaden16}. These are both so-called `red-back' systems, which harbour more massive companion stars than black-widow systems \citep{roberts13}, and the variations were accompanied by significant changes in the overall luminosity of the system.
In the case of \psr, however, the flux scaling factors in the $g$ and $i$-bands are consistent in the ULTRACAM and HiPERCAM data, to within the uncertainties in the flux calibration (see Section~\ref{sec:modelling}), a fact that is also evident from the direct comparison of the two light curves shown in Figure~\ref{fig:lc_asymm}. This implies that the luminosity at maximum remained approximately the same, implying a constant irradiation power in the two sets of observations and ruling out any significant variation in the pulsar emission or IBS, assuming that the latter mediates the irradiation as proposed by \cite{romani16}. Luminosity variations intrinsic to the companion star may also occur, for example due to the Applegate mechanism (\citealt{applegate92}, \citealt{applegate94}), but since the luminosity at light-curve maximum is dominated by the irradiation power this is unlikely to be detectable here. Besides, if intrinsic variations could in principle be detected at minimum luminosity, none have been reported in black-widow systems so far, to the best of our knowledge. Therefore, if one assumes that the extrinsic (irradiation) and intrinsic power sources remain stable, it seems plausible that the asymmetry observed in the ULTRACAM observations results from a variation of the heat redistribution pattern on the surface of the companion star, and the later HiPERCAM observations seem to indicate that this redistribution pattern alters on timescales of a decade or less. Repeat observations of the kind we present in this paper but with significantly shorter intervals between them may help to determine if the timescale on which the asymmetry appears is significantly shorter than this and whether or not it is periodic.

Finally, our assumption that the symmetric HiPERCAM light curves represent the underlying system provides us with a potential opportunity to compare and constrain the various asymmetric heating models, free of the degeneracies between the parameters that plagued the fits to the ULTRACAM data described at the end of Section~\ref{sec:modelling}. To this end, the posterior distributions of the fit parameters from the HiPERCAM photometry were modelled using a Gaussian mixture model and then used as priors for the fit parameters when fitting the ULTRACAM photometry (see \citealt{kennedy22}). We tried fitting both the spot model of \citet{clark21} and the convection model of \citet{voisin20}, but we were unable to obtain satisfactory fits with either model, particularly of the pre-maximum portion of the light curve. Either additional free parameters or a different parameterisation of the spot and convection models will be required to adequately model the ULTRACAM data.

\section{Conclusions}

Our light curves of \psr\ have demonstrated that whatever mechanism is responsible for the asymmetric heating observed on the companion stars in black-widow systems is likely to be variable on timescales of a decade or less and is most probably related to a change in the heat redistribution pattern on the stellar surface rather than any change in the irradiation power. We find that the companion star in \psr\ is close to filling its Roche lobe and has a mass and mean density consistent with a brown dwarf, but an apsidal motion constant that implies a significantly more centrally-condensed internal structure than is typical of such objects. We encourage continued monitoring of \psr\ to put tighter constraints on the heating-variability timescale, detailed modelling of the internal structure of the companion star to constrain the evolutionary history of the binary, and spectroscopy of the companion star, perhaps with JWST or the coming generation of extremely large telescopes, to measure the pulsar mass. 

\section*{Acknowledgements}

We thank the anonmyous referee for their comments on the manuscript. The design and construction of HiPERCAM was funded by the European Research Council under the European Union's Seventh Framework Programme (FP/2007-2013) under ERC-2013-ADG Grant Agreement no. 340040 (HiPERCAM). VSD, ULTRACAM and HiPERCAM operations are funded by the Science and Technology Facilities Council (grant ST/V000853/1). RPB, CJC, DMS, MRK and GV acknowledge support from the ERC under the European Union's Horizon 2020 research and innovation programme under Grant Agreement No. 715051 (Spiders). MRK acknowledges support from the Irish Research Council in the form of a Government of Ireland Postdoctoral Fellowship (GOIPD/2021/670: Invisible Monsters). DMS acknowledges the Fondo Europeo de Desarrollo Regional (FEDER) and the Canary Islands government for the financial support received in the form of grant number PROID2020010104. SGP acknowledges the support of a STFC Ernest Rutherford Fellowship. IP and TRM acknowledge support from the UK's Science and Technology Facilities Council (STFC), grant ST/T000406/1. The GTC is installed at the Spanish Observatorio del Roque de los Muchachos (ORM) of the Instituto de Astrof\'{i}sica de Canarias (IAC), on the island of La Palma. The WHT is operated on La Palma by the Isaac Newton Group in the Spanish ORM of the IAC. For the purpose of open access, the author has applied a Creative Commons Attribution (CC BY) licence to any Author Accepted Manuscript version arising.

\section*{Data availability}

The data presented in this paper will be shared on reasonable request to the corresponding author.

\bibliographystyle{mnras}
\bibliography{psr2051_refs}

\bsp
\label{lastpage}
\end{document}